\documentclass[conference]{IEEEtran}
\IEEEoverridecommandlockouts
\usepackage{cite}
\usepackage{amsmath,amssymb,amsfonts}
\usepackage{graphicx}
\usepackage{textcomp}
\usepackage{xcolor}

\usepackage{algorithm}
\usepackage{algpseudocode}
\usepackage{caption}
\usepackage{float}
\usepackage{listings}

\definecolor{codebg}{RGB}{245,247,250}
\definecolor{pykeyword}{RGB}{31,119,180}
\definecolor{pystring}{RGB}{214,39,40}
\definecolor{pycomment}{RGB}{44,160,44}
\definecolor{pyfunc}{RGB}{148,103,189}
\definecolor{pynumbers}{RGB}{110,110,110}

\lstdefinestyle{pythonstyle}{
    language=Python,
    backgroundcolor=\color{codebg},
    basicstyle=\ttfamily\footnotesize,
    keywordstyle=\color{pykeyword}\bfseries,
    stringstyle=\color{pystring},
    commentstyle=\color{pycomment}\itshape,
    identifierstyle=\color{black},
    emph={MapReduce,Job,build_containers},
    emphstyle=\color{pyfunc}\bfseries,
    numberstyle=\tiny\color{pynumbers},
    stepnumber=1,
    numbersep=6pt,
    frame=single,
    rulecolor=\color{black!15},
    showstringspaces=false,
    breaklines=true,
    tabsize=4
}

\lstdefinelanguage{json}{
    basicstyle=\ttfamily\footnotesize,
    numberstyle=\tiny\color{pynumbers},
    stepnumber=1,
    showstringspaces=false,
    breaklines=true,
    string=[s]{"}{"},
    comment=[l]{//},
    morecomment=[s]{/*}{*/},
    moredelim=[l][\color{pykeyword}]{:},
    morestring=[b]",
}

\lstdefinestyle{jsonstyle}{
    language=json,
    basicstyle=\ttfamily\footnotesize,
    stringstyle=\color{pystring},
    commentstyle=\color{pycomment},
    keywordstyle=\color{pykeyword},
    showstringspaces=false,
    breaklines=true
}

\def\BibTeX{{\rm B\kern-.05em{\sc i\kern-.025em b}\kern-.08em
    T\kern-.1667em\lower.7ex\hbox{E}\kern-.125emX}}
\begin{document}

\title{Design and Implementation of a Serverless MapReduce Framework for Scalable Data Pipelines\\}

\author{
\IEEEauthorblockN{1\textsuperscript{st} Angelos Dorotheos Chatzopoulos}
\IEEEauthorblockA{
\textit{Department of Informatics and Telecommunications}\\
\textit{National and Kapodistrian University of Athens}\\
Athens, Greece\\
angeloschatz@di.uoa.gr
}
\and
\IEEEauthorblockN{2\textsuperscript{nd} Babis Andreou}
\IEEEauthorblockA{
\textit{Department of Informatics and Telecommunications}\\
\textit{National and Kapodistrian University of Athens}\\
Athens, Greece\\
bcand@di.uoa.gr
}
\and
\IEEEauthorblockN{3\textsuperscript{rd} Kakia Panagidi}
\IEEEauthorblockA{
\textit{Department of Informatics and Telecommunications}\\
\textit{National and Kapodistrian University of Athens}\\
Athens, Greece\\
kakiap@di.uoa.gr
}
\and
\IEEEauthorblockN{4\textsuperscript{th} Stathes Hadjiefthymiades}
\IEEEauthorblockA{
\textit{Department of Informatics and Telecommunications}\\
\textit{National and Kapodistrian University of Athens}\\
Athens, Greece\\
shadj@di.uoa.gr
}
}

\maketitle

\begin{abstract}

Modern logistics systems tend to generate continuous streams of data from sources such as GPS, IoT sensors, and logistics management systems. The aggregation, processing, and analysis of data have become vital for monitoring operations, optimizing efficiency, and responding quickly to decision making tasks. In this paper, an event-driven MapReduce framework for real-time data processing in logistics environments is presented. This system runs on Kubernetes with Knative and utilizes Apache Kafka as the backbone for communication between the components. This platform is composed of five loosely coupled services that receive, process, and aggregate the incoming data in real-time. Redis is used to preserve workflow metadata, while an AWS S3 service provides persistent storage for the framework. The design is inspired by the MapReduce programming model. It integrates Function-as-a-Service (FaaS) principles with distributed processing techniques that allow configurable scaling based on the workload demands and the underlying hardware. Experimental evaluation shows that the system can scale effectively as the input data volume increases while supporting scale-to-zero, on-demand processing.

\end{abstract}


\section{Introduction}
\label{ch:introduction}

In recent years, applications have increasingly relied on continuous streams of data coming from sources such as tracking logs, warehouse events, and operational text records. The constant flow of information is a part of the everyday logistics operations, thus managing it effectively becomes a challenge. As data volumes continue to grow, traditional systems face increasing challenges related to scalability, management complexity, and cost efficiency. Frameworks such as Apache Hadoop \cite{hadoop} are designed to handle big data using the MapReduce model. They remove the need to manage low-level details such as scalability, fault tolerance, and availability. However, these frameworks usually rely on statically allocated clusters. This means that it can be expensive to run and may require skilled people to configure the system \cite{blendata_hadoop}.

Meanwhile, cloud computing has shifted towards more flexible and developer friendly execution environments. Serverless computing allows workloads to be split into smaller functions that scale and execute based on demand, eliminating the requirement to have an already configured infrastructure. The cloud platform automatically manages scaling, execution, and resource provisioning \cite{bashir2018serverless}. The cloud provider bills only for the resources used and the time the application code is running \cite{lin}. This makes it an optimal architecture to run workloads often found in real-time systems.

However, combining traditional data processing methods like MapReduce with a serverless environment introduces several challenges. The ephemeral nature of functions, cold starts, short execution lifetimes, and limited data locality complicate tasks such as data storage and component coordination \cite{hassan}. Existing serverless environments are mainly designed for event-driven or request-based architectures rather than large scale batch processing for real-time data.

This paper focuses on designing a serverless, event-driven architecture for real-time data, that applies a MapReduce processing model. The goal is to create and deploy a framework that can effortlessly process, scale, and manage large-scale data workloads. It is organized as follows: Section \ref{ch:background} showcases related works, the background and the contributions of this research. Section \ref{ch:design} breaks down the proposed solution. The evaluation of the proposed solution and the experiments performed are detailed in Section \ref{ch:results}. Finally, the conclusions are presented in Section \ref{ch:conclusion}.

\section{Background and Related Work}
\label{ch:background}

This section presents the core theory necessary to understand the proposed framework, including the prior work and the main objective of this paper.

\subsection{Background}

\subsubsection{MapReduce Model}

MapReduce is a programming paradigm designed to process large volumes of data concurrently to improve performance. 
It is highly scalable, since data can be divided into smaller chunks that can be processed in parallel. MapReduce consists of two main stages represented by the following functions: \textit{map}, which processes input data and produces intermediate key-value pairs (records), and \textit{reduce}, which aggregates the intermediate results to generate the final output. The model is inspired by the \texttt{map} and \texttt{reduce} functions used in functional programming. 

\textit{MapReduce} was originally introduced by Google \cite{google-mapreduce} and has since become a general term for data processing computation methodology. Between the  \textit{map} stage and the  \textit{reduce} stage, there is a phase called \textit{shuffle}. Shuffling is a necessary stage to group all the values corresponding to each key and assign them to a single reducer so that there are no partially aggregated results calculated. Figure \ref{fig:mapreduce_types} shows the functional signatures of the \texttt{map} and \texttt{reduce} operations, displaying how they convert the data between stages of the MapReduce algorithm. 

In some cases, map functions produce a large volume of intermediate data. This data must be stored in persistent storage, so that the corresponding reducers can retrieve the appropriate values for each key and perform the reduction. To reduce network congestion, each Mapper can run a local reduce function called a \textit{combiner}, before writing the records to persistent storage. This decreases the amount of data written and reduces the volume of input that each reducer must merge and read.

\begin{figure}[H]
\centering
\captionsetup{justification=centering}
\[
\textbf{map}\ (k_1, v_1) \rightarrow \text{ list}(k_2, v_2)
\]
\[
\textbf{reduce}\ (k_2, \text{list}(v_2)) \rightarrow \text{ list}(v_2)
\]
\caption{Functional representation of Map and Reduce operations.}
\label{fig:mapreduce_types}
\end{figure}

\subsubsection{Serverless Computing Model}

Serverless computing is a cloud computing paradigm that abstracts server management to a cloud provider. This lets developers concentrate on writing and deploying application logic as functions \cite{bashir2018serverless}. The term \textit{Serverless} does not mean that there are no servers involved. Instead, the cloud provider dynamically allocates resources, executes the application code, and handles scaling based on specific configurable metrics. Programmers focus on developing the application without controlling the underlying infrastructure. 

The main benefits \cite{bashir2018serverless} of serverless computing are the pay-per-use model, where users are billed only based on the execution time and resources their functions consume. Moreover, it provides flexibility towards the programming language that can be used to implement the functions. Lastly, a serverless architecture can scale based on the workload demands or incoming traffic automatically based on the users' configuration.

However, a serverless architecture poses several challenges as well. Cold starts may impose additional latency overhead due to the initialization of a service. Additionally, triggered functions do not provide any persistent state, making it more difficult to store and fetch data. Moreover, Cloud-native functions such as AWS Lambda and Google Cloud Functions have a defined maximum execution time, while the underlying hardware is abstracted from the developer, leading to a lack of fine-grained optimizations. Finally, serverless environments may introduce several security challenges related to multi-tenancy, function isolation, and dependency management.

\subsubsection{Kubernetes}

Kubernetes \cite{kubernetes_docs} is an open-source orchestration platform that automates the deployment, scaling, and management of containers based on declarative configurations. It is a framework designed to run distributed services, and handle the scaling and failover of them. Containers are executable application components that package the application code along with the required dependencies to be able to run the code in multiple environments.

\subsubsection{Knative}

Knative \cite{knativeDocs} is a Kubernetes-based platform that provides a set of middleware resources for building, deploying, and managing serverless workloads. Knative provides a high level abstraction that accelerates the development productivity of cloud-native serverless applications. This allows containers to be created on-demand, scaling from zero to any required level configured by a developer. Knative offers a streamlined and standardized developer experience for building and deploying stateless functions, without the need to deal with multiple steps and tools to run the application on Kubernetes.

\subsubsection{Apache Kafka}
\label{sec:kafka}

Apache Kafka \cite{kafkaIntro} is a distributed event streaming platform that can be deployed in multiple environments. Event records consist of a key, value, timestamp, and optional metadata headers. Producer applications publish events to Kafka, while the consumers are those that subscribe to these events. Events are stored in topics. A topic can have multiple producers and consumers, and is divided into partitions that can be located on different Kafka brokers.

\subsection{Related Work}
\label{ch:related-work}

In recent years, the focus on MapReduce systems has shifted from traditional cluster based architectures to serverless FaaS models. Prior work on both architectures are presented, to highlight the distinctions from our choices.

\subsubsection{Traditional MapReduce Systems}

The traditional MapReduce model was introduced and popularized by Google \cite{google-mapreduce} and later implemented in an open-source framework Hadoop \cite{hadoop}. These systems are typically designed to run on persistent clusters, which limits operational costs and flexibility, and ease of use. Moreover, Apache Spark \cite{spark} is an in-memory processing model with a Directed Acyclic Graph (DAG) engine that offers improved performance for many workloads and provides a more user-friendly API. However, it still generally requires running on dedicated clusters, which means that users must maintain the nodes and storage systems, without natively providing on-demand processing.

Unlike cluster-based systems, the design described in this paper relies on short-lived containers running on Kubernetes nodes and scales based on the number of Kubernetes worker nodes that the user configures.

\subsubsection{Serverless MapReduce Systems}

Multiple serverless approaches have been proposed in recent years. IBM-PyWren \cite{ibm-pywren} demonstrates a serverless architecture using IBM Cloud Functions (based on Apache OpenWhisk) with IBM Cloud Object Storage (IBM COS) to store the data and intermediate results. IBM Cloud Functions runtime is based on Docker images, which, compared to other MapReduce frameworks, allows to customize Docker images and to include additional required packages for running user-defined functions.

A similar approach was taken in recent studies and system designs such as by Xu Huang et al., BabelMR, MARLA, StarData, Corral \cite{xu-huang,babelmr,marla,stardata,congdon2018corral}, where the map and reduce stages are executed using the AWS Lambda compute service, with S3 serving as the persistent storage for intermediate and final results. These systems introduce several partitioned data shuffling methodologies, and components' scheduling policies to achieve their goal. 

Finally, Jingwen Cai \cite{jingwen} et al. presented a serverless MapReduce workflow running on Alibaba Cloud, using Function Compute Service (FC) for stateless mappers and reducers, and Alibaba Cloud Object Storage Service (OSS) for storage. They show that function concurrency and CPU capacity significantly affect the performance of data processing, while memory allocation has minimal impact on workers' response time.

These approaches show that designing an embarrassingly parallel data processing task, such as MapReduce, in a serverless architecture is a promising way to exploit massive concurrency with minimal operational effort. However, these systems are constrained by the side effects of the functional runtimes. Most FaaS frameworks allow only limited memory, storage, and execution time. In addition, they heavily rely on platform-specific cloud services, which reduces portability since they are tightly coupled to a specific cloud provider \cite{patsch}.

\subsection{Contributions}

This work presents a serverless, event-driven data processing framework inspired by the MapReduce model. The primary goal of this paper is to design and implement a system that allows parallel execution of data processing tasks by dividing the workflow into independent stages. This allows the users to easily configure and submit data processing workflows without the need to handle the underlying infrastructure or deployment details. The framework is built to support configurable pipelines where each processing stage can be individually scaled based on the workload demands.

\section{System design}
\label{ch:design}

This section presents the internal design of the MapReduce framework. The design of the framework utilizes the technologies discussed in Section \ref{ch:background}, and the components are implemented using Python, with all components packaged as containerized services.

\subsection{Framework Design Architecture}

The MapReduce system design consists of a coordinator-workers architecture. The Coordinator controls the execution of the MapReduce workflow while worker nodes are triggered (through Kafka broker) and created  to process the assigned data. The Coordinator component is a Knative Service that is triggered using an HTTP request from the client, while the worker components are Knative JobSinks that the Coordinator creates by producing CloudEvents \cite{knativeDocs}. Worker components can communicate with a Coordinator service using an HTTP request. All components are able to scale from zero replicas.

\subsubsection{Coordinator component}

The Coordinator component is responsible for managing the execution of each MapReduce job. Firstly, it receives a request from the client with a specific JSON configuration format, as it is the entry point of the system. It assigns work to the Splitter, creates and synchronizes the Mapper, Reducer, and Finalizer containers. It sends them the necessary data through the Kafka broker. The Coordinator also receives updates from the created components to track the progress of the MapReduce workflow. The Coordinator is the entry point of a MapReduce task. In case of any failure, it updates the job state progress metadata in Redis. Multiple workflows can be managed from a single Coordinator since it is stateless.

\subsubsection{Splitter component}

The Splitter is responsible for dividing the data into chunks before the Mappers begin. Given an input (S3 path prefixes), it measures the total size of the data and splits it according to the number of Mapper containers that will be created (based on user configuration) to equally distribute the payload. These chunks are uploaded to Redis as byte ranges metadata, so each Mapper can fetch the assigned data from the S3 bucket. In case of the input being text-based, the splitter extends the boundaries it will split, in order to not cut any record in half. Otherwise, records are split solely based on the byte offsets.

\subsubsection{Mapper component}

Mapper components process each input chunk that is stored as metadata in Redis and fetches them from the S3. Then, it starts executing the user-defined function to produce intermediate key-value pairs. These intermediate key-value pairs are stored as spill files into an S3 bucket. The results are partitioned based on a hash function over the key which outputs the target Reducer that this record is going to be processed. Each spill file, before being uploaded, is sorted (grouped) by key, so the Reducer only needs to find the assigned spill files and merge them, thus contributing to the shuffle phase.  Each mapper component runs in parallel as a stateless Knative Jobsink. Lastly, when the execution of a Mapper is completed, it notifies the Coordinator that its assigned task has finished.

\subsubsection{Reducer component}

The Reducer fetches the intermediate data produced by the Mappers and creates a single merged file where records are grouped by key. This is required so all values linked to a key are processed together. Each Reducer reads the Mappers' spill files corresponding to the reducer ID, based on the name of the spill file (format \texttt{spill-{reducer\_id}-{file\_index}-{mapper\_id}}). It retrieves its corresponding intermediate spill files from S3 and runs the k-way merge algorithm (\textit{k} configured by the user in the JSON file). Then, each Reducer spawned applies the user-defined function to produce the final aggregated results. Merging is performed so that, for each key, all values are processed before moving to the next. Each Reducer creates a single output file. Similar to the Mapper, when the execution is completed, it notifies the Coordinator that its assigned task has finished.

\subsubsection{Finalizer component}

The Finalizer component is a single spawned component that collects the output files produced by the Reducers and combines them into a single output file. Since S3 does not support updates on the same file, the Finalizer needs to stream the contents of each Reducer output into a single object file.

\subsection{Storage Layer}

The system uses a shared storage layer based on S3 and Redis. The S3 is used to store input files, intermediate Mapper files and final results, while the metadata are saved using Redis. The user has to specify the location into an S3 bucket the input data are located in. Similarly, the final output results are stored into the S3 bucket. 

\begin{figure}[h!]
    \centering
    \includegraphics[width=0.60\linewidth]{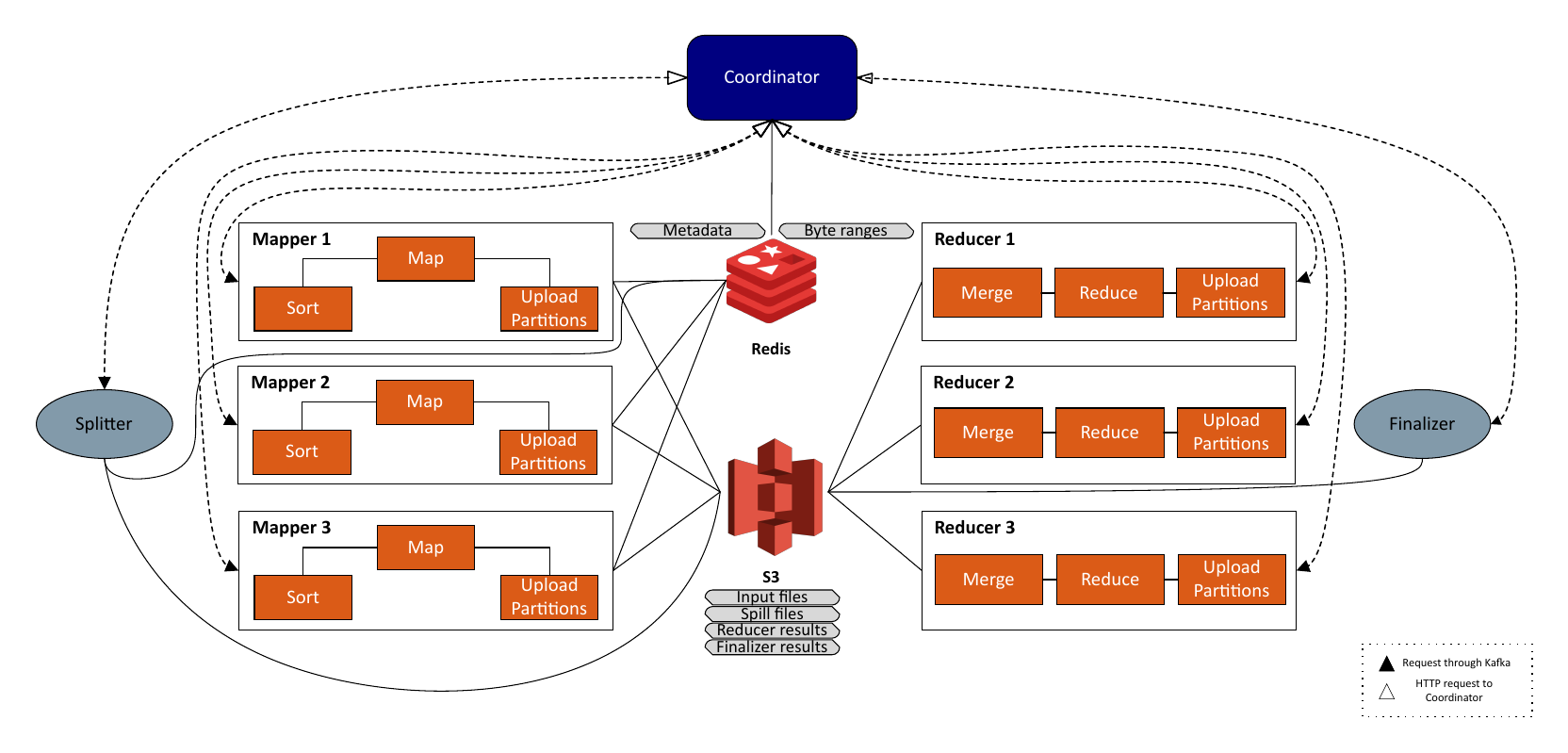}
    \caption{Overall architecture of the serverless MapReduce inside Kubernetes.}
    \label{fig:mapreduce-arch}
\end{figure}

Figure \ref{fig:mapreduce-arch} shows the overall system architecture. It is noted that Reducer and Finalizer components are optional. The workflow can be used to execute only a Mapping phase. Lastly, it is not required the number of Mapper components to be equal to the number of Reducer components.

\subsection{Configuration Input of a MapReduce Job}
\label{sec:input-mapreduce}

The input for a MapReduce job is provided as a JSON file by the user, containing the necessary information to execute the specific MapReduce task. This file defines the input and output locations in S3, component parameters, and the overall structure of the workflow in a simple structure. The configuration includes the number of Mapper and Reducer components, optional execution of the Finalizer, split boundaries, and whether the data should be handled as binary. Additionally, it configures the input and output buffer sizes, the threshold as a percentage required to upload data to persistent storage, and includes the source code defined by the user for the Map and Reduce functions.

\subsection{Client Interaction}

A user interacts with the MapReduce system using a Python package. The package  builds the container images, including Redis, uploads them to a container registry, and deploys them to Kubernetes.

Before running a task, the user defines the MapReduce functions (\texttt{mapper} and \texttt{reducer}) in Python and configures the input JSON file specifying the dataset and job parameters. A \texttt{yaml} file is also used to configure system-level settings such as S3 settings, Kafka broker URL, Redis host and port, and the Coordinator host and port, depending on the Kubernetes deployment environment. In case additional libraries are required, the user can define them in a requirements file on the Mapper or Reducer component and rebuild the containers.

Once the configuration is ready, and image building and deployment are complete, the user can submit and execute MapReduce jobs through the Python package. The package sends the JSON input, including the user-defined MapReduce functions, to the Coordinator via request to start the worfklow.

\begin{figure}[h!]
    \centering
    \includegraphics[width=0.7\linewidth]{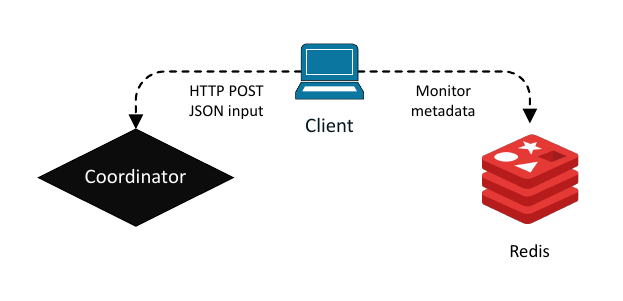}
    \caption{Client interaction for submitting and monitoring MapReduce jobs.}
    \label{fig:client-system}
\end{figure}

The package then monitors the job's progress by querying the Redis metadata, tracking whether the task is still running, has finished successfully, or has failed. Figure \ref{fig:client-system} shows the interaction the Python package performs underneath, to submit and monitor a new MapReduce task.

Figure \ref{fig:mapreduce-client-example} shows an example of how a user can define two jobs to run in parallel. First, container images must be built before running any MapReduce job. The first job will run the \texttt{mapper\_fn} and \texttt{reducer\_fn} functions, while the second job will run the \texttt{mapper\_fn2}, \texttt{mapper\_fn3}, and \texttt{reducer\_fn2} functions. The functions must be defined by the user so that the Python package can extract the source code, append it to the JSON payload, and send the request to the Coordinator to begin the MapReduce workflow. In this example, the first job runs one map and one reduce function, while the second job runs two map functions followed by a reduce function.

In reality, the second job, which has two map functions and one reduce function, is executed as two distinct MapReduce jobs: the first executes the first map function only, and the second executes the second map function and the final reduce function. However, the user only needs to configure the initial JSON payload. The Python package handles locating the intermediate files from the first map function to continue with the second map function. The configuration can be loaded from a JSON file, and function definitions do not need to be included in the configuration, since the package fetches them and appends them to the JSON payload before sending the request to the Coordinator.

The package is designed to execute each job as an asynchronous Python operation, because multiple map functions and different MapReduce jobs can be run in parallel before reduction. Each operation sends the configuration payload to the Coordinator and awaits completion or failure. Once the MapReduce jobs have finished, the package returns the job ID for each job, allowing users to identify and inspect the results in S3 storage. Figure \ref{fig:mapreduce-functions} shows an example of word counting functions that can be passed to the Python package to begin MapReduce processing.

\begin{figure}[h!]
\centering

\begin{lstlisting}[style=pythonstyle]
build_containers()

config1 = load_json("config_job1.json")
config2 = load_json("config_job2.json")

job_list = [
    Job(
        payload=config1,
        mappers=[mapper_fn],
        reducer=reducer_fn
    ),
    Job(
        payload=config2,
        mappers=[mapper_fn2, mapper_fn3],
        reducer=reducer_fn2
    )
]

mapreduce = MapReduce(
    coordinator=coord_service,
    jobs=job_list,
    redis_host=redis_host,
    redis_port=redis_port,
    logging=True
)

job_results = await mapreduce.run()

print("Completed jobs:", job_results)
\end{lstlisting}

\caption{Example of submitting and executing MapReduce jobs using the framework's client package.}
\label{fig:mapreduce-client-example}
\end{figure}

\begin{figure}[h!]
\centering

\begin{lstlisting}[style=pythonstyle]
def mapper(key, chunk):
    for word in chunk.split():
        yield word, 1

def reducer(key, values):
    total = sum(values)
    return key, total
\end{lstlisting}

\caption{Example Map and Reduce functions used in the framework.}
\label{fig:mapreduce-functions}
\end{figure}

\section{Results}
\label{ch:results}

This section evaluates the performance of the MapReduce implementation across a range of input data sizes by counting words using the map and reduce functions depicted in Figure \ref{fig:mapreduce-functions}. The performance for each phase (processing, uploading, and downloading) is presented for each component, as well as the total execution time on a fixed number of Mapper and Reducer components.

\subsection{Deployment}

The framework was deployed on Amazon Elastic Kubernetes Service (EKS), which provides a managed Kubernetes environment in the cloud. Five worker nodes of type \texttt{c4.2xlarge} where used to run the experimental evaluation.

The Coordinator was deployed with a fixed pod count of one and did not scale from zero. Since HTTP requests required to be port forwarded from the Python client, it was necessary to be alive. In contrast, all other stateless worker components, were scaled from zero instances and automatically instantiate based on the workload's configurations.

\subsection{Input Data}

The input data comes from a Wikipedia dataset in English, obtained through HuggingFace \cite{wikidump}. To improve the locality of word occurrences in the input, a light preprocessing stage is applied. All words are converted to lowercase, punctuation is removed and any additional whitespace is deleted.

Since the input is text based, improving locality increases the probability that identical words appear close together in each Mapper input buffer. This improves the effectiveness of the combiner and reduces the volume of intermediate data that must be uploaded and later merged by Reducers.

\subsection{Performance Results on AWS EKS}

The experimental results running the MapReduce system are presented with a specific client configuration: combiner and Finalizer enabled, buffer input size and buffer output size set to 50MB, multipart size set to 5MB, Reducer merge size set to 100, and buffer threshold set to 75\%.

\begin{figure}[h!]
\centering
\includegraphics[width=0.75\linewidth]{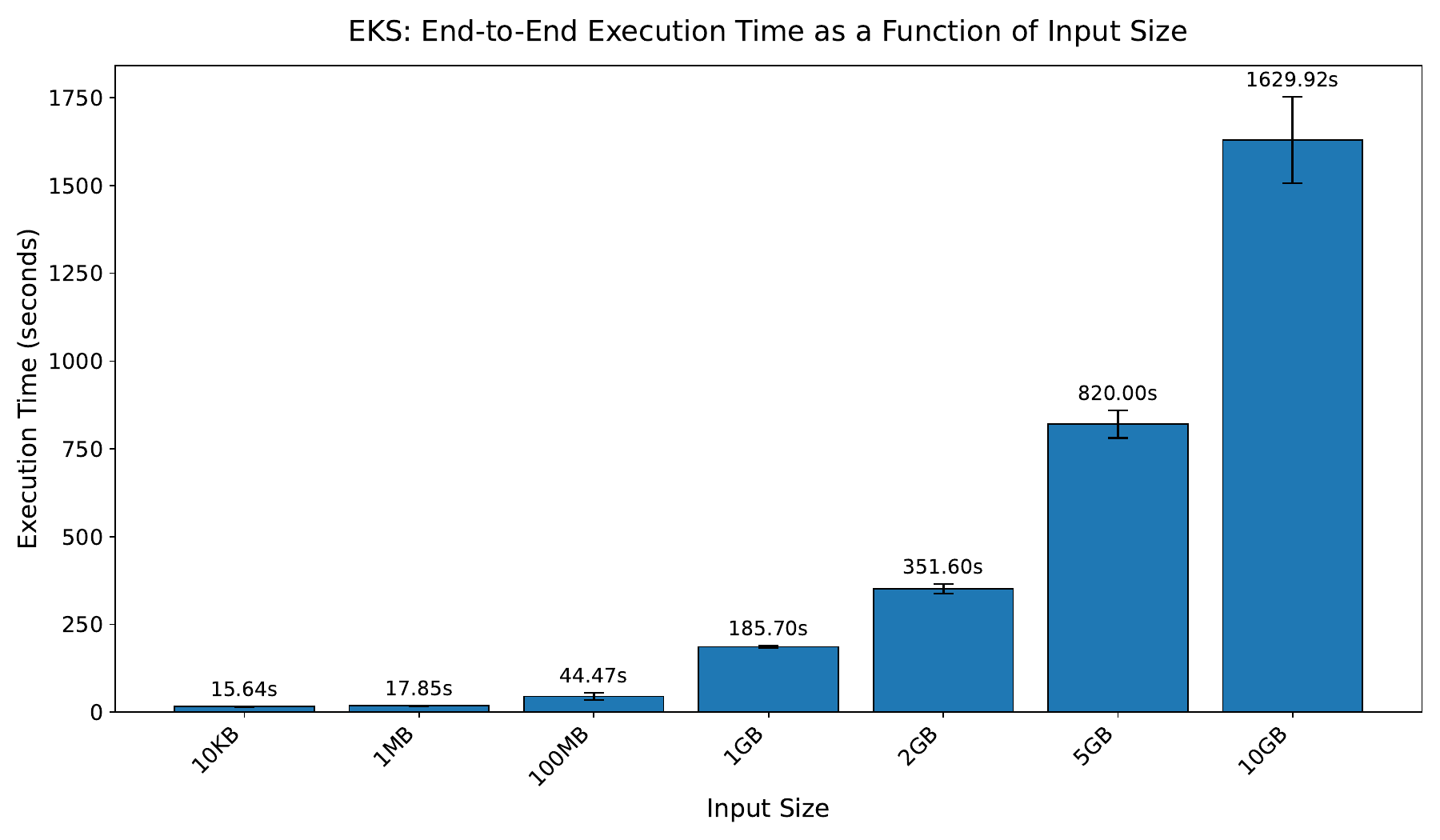}
\caption{End-to-end MapReduce execution time on EKS for increasing input sizes with 4 Mappers and 2 Reducers.}
\label{fig:job-durations}
\end{figure}

The experimental evaluation was done with a small fixed number of Mappers and Reducers: four Mappers and two Reducers. The execution time is measured as the input size grows. Figure~\ref{fig:job-durations} shows the end to end execution time (from the start to the completion of each MapReduce workflow) as the input size increases. As the input grows, a roughly linear trend in total execution time is observed. However, on very small data volumes, performance does not scale linearly. This is due to the cold starts of the components dominating and outweighing the actual processing time.

\begin{figure}[h!]
\centering
\includegraphics[width=0.8\linewidth]{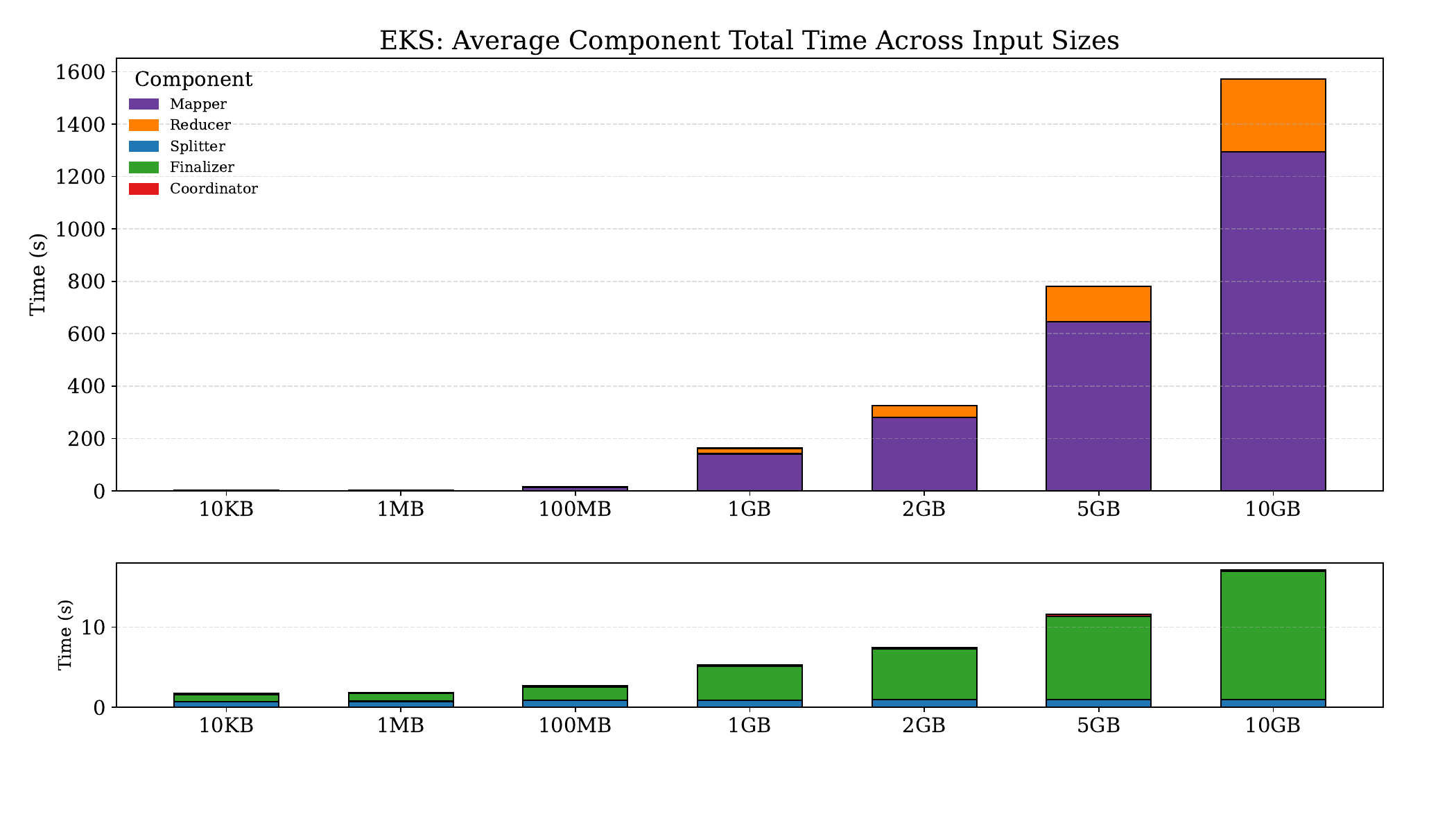}
\caption{Average total time of each MapReduce component on EKS for increasing input sizes with 4 Mappers and 2 Reducers.}
\label{fig:stacked-components}
\end{figure}

Figure~\ref{fig:stacked-components} presents the average total time per component across the input sizes using two Mappers and Reducers. The Mapper contributes the most execution time, since it must sort the buffer and run the combiner before uploading data to S3. Overall, the Coordinator, Splitter, and Finalizer introduce only a very small overhead as the input size grows, leaving the main costs to Mappers and Reducers, which allows the user to scale them as needed.

\begin{figure}[h!]
\centering
\includegraphics[width=0.9\linewidth]{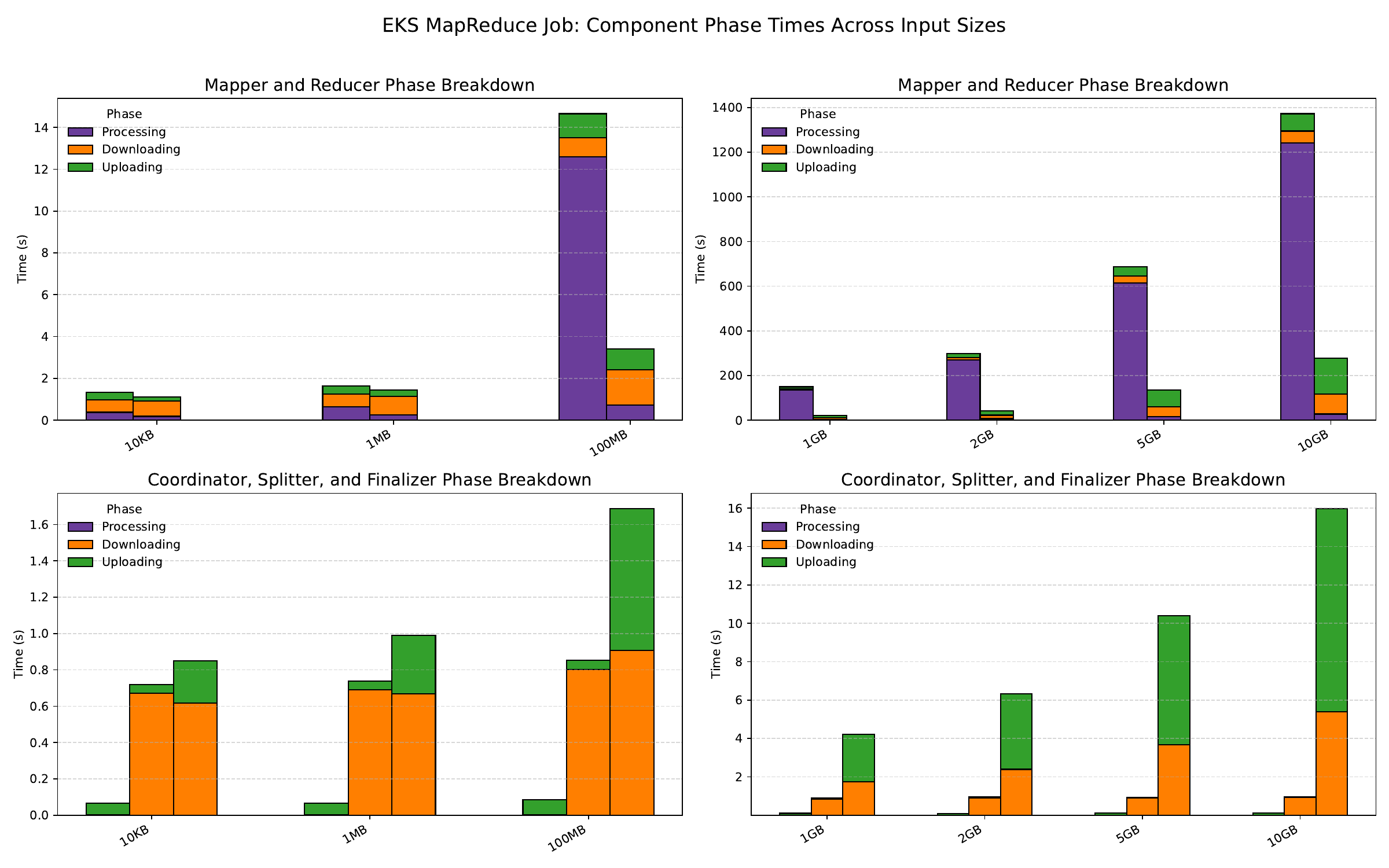}
\caption{Average total time of each component across all phases (processing, uploading, downloading) on EKS for increasing input sizes with 4 Mappers and 2 Reducers.}
\label{fig:all-inputs-phase-comps}
\end{figure}

Figure~\ref{fig:all-inputs-phase-comps} shows, for each component, the stacked execution time across all phases as the input size increases. The Mapper dominates execution time due to buffer management, sorting, and running the combiner before each upload once the buffer reaches a threshold (processing time). The Reducer, on the other hand, only needs to merge intermediate files, run the reduction function, and upload the final output into a single file.

\section{Conclusion}
\label{ch:conclusion}

This paper presents the design and implementation of a serverless, event-driven MapReduce framework. The main objective was to design, build, and evaluate a system capable of handling large volumes of data. The framework provides a task-per-service model, where each service executes independently. This offers users a straightforward way to run data processing tasks effortlessly. From a logistics viewpoint, this is relevant for processing continuous streams of data from distributed systems.

The strengths and limitations of serverless architectures were examined, with a particular focus on how these constraints affect data-processing workloads. These observations motivated the design of a new platform that avoids monolithic execution by splitting the workflow into loosely coupled components. The system was deployed and evaluated on Amazon EKS. The evaluation shows that simple serverless functions can efficiently support embarrassingly parallel data-processing tasks, such as the MapReduce model.

\bibliographystyle{IEEEtran}
\bibliography{references}

\end{document}